\documentclass[12pt]{iopart}

\usepackage{graphicx}

\newcommand{\fields}{\phi_{i_{1}}...\phi_{i_{p}} }
\newcommand{\fieldls}{\phi_{i_{1}}...\phi_{i_{p-1}} }
\newcommand{\couple}{J_{i_{1}...i_{p}}}
\newcommand{\couplei}{J_{i,i_{1}...i_{p-1}}}
\newcommand{\invcouple}{K_{i_{1}...i_{p}} }
\newcommand{\sumN}{\sum_{i=1}^{N} }
\newcommand{\sump}{\sum_{i_{1}...i_{p}} }
\newcommand{\znoise}{\langle Z[\{l_{i}\},\{h_{i}\}] \rangle}
\newcommand{\z}{\overline{\langle Z[\{l_{i}\},\{h_{i}\}] \rangle}}
\newcommand{\zJ}{\overline{\exp(L_{J}[\phi,\hat{\phi}])}}

\newcommand{\hfield}{\hat{\phi}}
\newcommand{\vph}{\varphi}
\newcommand{\hvp}{\hat{\varphi}}
\newcommand{\intfin}{\int_{t_{0}}^{t_{f}}}
\newcommand{\ar}{\alpha_{i_r}}
\newcommand{\ai}{\alpha_{i}}
\newcommand{\p}{\phi^{0}}
\newcommand{\np}{{\phi'}^{0}}
\newcommand{\pp}{\phi^{1}}
\newcommand{\npp}{{\phi'}^{1}}

\newcommand{\intgg}{\int\limits_{0\leq t_1,t_2<\infty \atop \tau_{0}\leq|t_1-t_2|}}
\newcommand{\intft}{\int\limits_{\tau_{0}\leq|t_1-t_2|<b\tau_{0} \atop 0\leq t_1,t_2<\infty}}

\newcommand{\q}{\psi^{0}}
\newcommand{\qq}{\psi^{1}}

\begin{document}
\title{Time reparametrization invariance in arbitrary range p-spin
  models: symmetric versus non-symmetric dynamics} 
\author{Gcina~A.~Mavimbela and Horacio~E.~Castillo}
\address{Department of Physics and Astronomy, Ohio University, Athens,
  Ohio, USA, 45701} 
\ead{\mailto{mavimbel@helios.phy.ohiou.edu}, \mailto{castillh@ohio.edu}}

\begin{abstract}
We explore the existence of time reparametrization symmetry in p-spin
models. Using the Martin-Siggia-Rose generating functional, we
analytically probe the long-time dynamics. We perform a
renormalization group analysis where we systematically integrate over
short timescale fluctuations. We find three families of stable fixed
points and study the symmetry of those fixed points with respect to
time reparametrizations. One of those families is composed entirely of
symmetric fixed points, which are associated with the low temperature
dynamics. The other two families are composed entirely of
non-symmetric fixed points. One of these two non-symmetric families
corresponds to the high temperature dynamics.

Time reparametrization symmetry is a continuous symmetry that is
spontaneously broken in the glass state and we argue that this gives
rise to the presence of Goldstone modes. We expect the Goldstone modes
to determine the properties of fluctuations in the glass state, in
particular predicting the presence of dynamical heterogeneity.
\end{abstract}

%
%
%

\date{\today}

\maketitle
\section{Introduction}
Very slow dynamics is an essential feature of glasses~\cite{deben}. In
both structural glasses and spin glasses slow dynamics is manifested
through a dramatic increase in relaxation times. This slowdown has
been captured in mean field theories, such as the mode coupling theory
for supercooled liquids~\cite{MCT} and the dynamical theory for mean
field spin glass models~\cite{somp, bouch,
  Cugliandolo-Kurchan-jphysa94, Cugliandolo-Kurchan-prl-93}. Even
though mean field theories are useful in describing some aspects of
glassy dynamics, they do not completely capture phenomena associated
with fluctuations. Fluctuations have been shown, particularly with the
discovery of {\em dynamical heterogeneities}, to be central to an
understanding of glassy dynamics~\cite{Ediger-arpc-2000}.

Dynamical heterogeneities - mesoscopic regions that evolve differently
from each other as well as from the bulk - have been found in
experimental studies of materials close to the glass
transition~\cite{wks1,wks2, Courtland-Weeks-jphys-03} and in
simulations of both spin glasses and structural glasses~\cite{Pari,
  Kob,glotz}.  Their presence has been directly observed at the
microscopic level in experiments on colloidal
glasses~\cite{Courtland-Weeks-jphys-03} and granular
systems~\cite{Durian}. Understanding the onset of heterogeneities
without an apparent structural trigger is believed to be key to an
understanding of the glass transition~\cite{Ediger-arpc-2000}.  There
have been several theoretical attempts to explain the emergence of
heterogeneous dynamics as the glass transition is approached. One of
them is a geometrical picture, according to which dynamical
heterogeneities result from non-trivial structure in the space of
trajectories due to dynamical constraints~\cite{garra}.  Another
proposed explanation is provided by the Random First Order transition
(RFOT) approach, in which a liquid freezes into a mosaic of aperiodic
crystals~\cite{woly}.  Here we will explore a different theoretical
avenue to explain dynamical heterogeneities, which is based on time
reparametrization symmetry~\cite{ckcc-prl-2002, cccik-prb-03,
  Castillo-prb-08}.

Time reparametrization symmetry (TRS), the invariance under
transformations of the time variable $t\rightarrow h(t)$, was
discovered some years ago in the mean-field non-equilibrium dynamics
of the Sherrington-Kirkpatrick model and the p-spin model~\cite
{Cugliandolo-Kurchan-jphysa94, Cugliandolo-Kurchan-prl-93}. The
symmetry, which was shown to be present in the long-time limit of the
mean field evolution, implies that the asymptotic equations do not
have a unique solution~\cite{Cugliandolo-Kurchan-jphysa94,
  Cugliandolo-Kurchan-prl-93, Cugliandolo-arxiv-2002}.  In more recent
studies, TRS has been proved to be present in the long time dynamics
of the glass state in a short range spin glass model, the
Edwards-Anderson model~\cite{ckcc-prl-2002, cccik-prb-03,
  Castillo-prb-08}. In this last case, the proof of the symmetry is at
the level of the generating functional, including all fluctuations.
Using the renormalization group (RG), it was shown that the stable
fixed point of the generating functional corresponding to glassy
dynamics is invariant under reparametrizations of the time
variable. However not all models of interacting spins under Langevin
dynamics show this behavior. For example, in a study of the O(N) model
it was shown that the symmetry is not present, even for the long time
limit of the low temperature dynamics~\cite{chamon}. The explanation
for dynamical heterogeneities from TRS is derived from the fact that
TRS is spontaneously broken by the correlations and responses in the
glass state. A spontaneously broken continuous symmetry is expected to
give rise to Goldstone modes, and these modes are associated with
spatially correlated fluctuations of the time variable, which give
rise to heterogeneous dynamics. The proposal that dynamical
heterogeneities originate in reparametrization fluctuations of the
time variable is supported by positive evidence from numerical studies
in spin glasses~\cite{cccik-prb-03} and in structural
glasses~\cite{Castillo-Parsaeian-nphys2007,
  Avila-Parsaeian-Castillo-arxiv-2010}.

In the present work we go beyond mean field theory and use a
renormalization group procedure to study the presence of time
reparametrization symmetry in the long time dynamics of p-spin models
with arbitrary interaction range, including all fluctuations. We
consider a system of soft spins on a lattice, with p-spin
interactions. The spin couplings are assumed to be uncorrelated
Gaussian random variables with zero mean. We assume Langevin dynamics
for the spins with a white noise term that represents the coupling of
the spins to a heat reservoir.  We set up the calculation by writing
the generating functional of the spin correlations and responses using
the Martin-Siggia-Rose (MSR) approach and introduce two-time fields
that are associated with the spin correlations and responses. To study
the long time dynamics we start the renormalization group procedure by
introducing a short time cutoff $\tau_{0}$.  We systematically
increase the short time cutoff by integrating over the two-time fields
associated with the shortest time differences, thus following a
procedure analogous to Wilson's approach to the RG.  In our case,
however, we integrate over fluctuations that are fast in {\em time\/},
not in space.  We find three families of stable fixed points. The
first family corresponds to fixed point actions containing the
coupling to the thermal bath but not the spin-spin interactions. The
fixed points in this family are not time reparametrization invariant,
and we believe that this family is associated with the high
temperature dynamics. A second family of stable fixed points that are
not time reparametrization invariant corresponds to fixed point
actions containing both the coupling to the thermal bath and the
spin-spin interactions. For the third family, the spin-spin
interaction term is marginal but the coupling to the thermal bath is
irrelevant. The fixed points in this last family are time
reparametrization invariant, and we believe that they represent the
low temperature glassy dynamics of the model. After obtaining these
results, we discuss their connection with dynamical heterogeneity in
the p-spin model, and we speculate on how a similar procedure may be
applied to models of structural glasses, which have been shown to be
connected to the p-spin model~\cite{Kirkpatrick-etal,
  Moore-Drossel-prl-2002}.

The rest of the paper is organized as follows: in Sec.~\ref{sec:model}
we start by giving a description of the model and an illustration of
how we derive the Martin-Siggia-Rose generating functional; in
Sec.~\ref{sec:RG} we show how we use Wilson's approach to the
renormalization group to get stable fixed points; in
Sec.~\ref{sec:TRS} we study the stable fixed point generating
functionals and determine which ones are invariant under
reparametrizations of the time variable; and in Sec.~\ref{sec:conc} we
end with a discussion of our results and conclusions.

\section{Model and MSR generating functional}
\label{sec:model}
The p-spin Hamiltonian is given by
  \begin{equation}
   H=-\frac{1}{p!}\sump \couple \fields,
  \end{equation}  
where the $\{\phi\}_{i=1,\cdots, N}$ are soft spins subject to the
spherical constraint $\sumN [\phi_{i}(t)]^{2}=N$, and the couplings
are assumed to be uncorrelated, Gaussian distributed, zero mean random
variables, $P\{J\}=\prod_{i_{1}<...<i_{p}}
\frac{1}{\sqrt{4\pi\invcouple}} \exp[-\couple^{2}/4\invcouple]$.  The
dynamics is given by the Langevin equation
\begin{equation}
  \Gamma_{0}^{-1}\partial_{t}\phi_{i}(t)= -\frac{\delta
    H}{\delta\phi_{i}(t)}+\eta_{i}(t), 
\end{equation}
where the $\{\eta_{i}(t)\}_{i=1,...,N}$ are assumed to be zero mean
gaussian random variables with the correlation
$\langle\eta_{i}(t)\eta_{j}(t')\rangle=2T\delta_{ij}\delta(t-t')$ that
couple the spins to a thermal bath at temperature $T$. Then the
Langevin equation can be written as
\begin{equation}
  \Gamma_{0}^{-1}\partial_{t}\phi_{i}(t) =
  \frac{p}{p!}\sum_{i_{1}...i_{p-1}}\couplei\fieldls+\eta_{i}(t). 
\end{equation}

From the Langevin equation we use the Martin-Siggia-Rose
formalism~\cite{DeDominicis-Peliti-prb-78} and write down the noise
averaged generating functional
\begin{eqnarray}
  \znoise=\int D\phi D\hfield D\hvp D\hat{N}
  \exp\left\{L[\phi,\hfield]+
  \sumN\int_{t_{0}}^{t_f}dt [l_i(t)\phi_i(t)\right. \nonumber \\* 
  \left.+ih_i(t)\hat{\phi}_i(t)] +i\sumN\hvp_{i}[\phi_{i}(t_{0})-
    \vph_{i}]+
  i\int_{t_{0}}^{t_{f}}dt \hat{N}(t)
  \left[\sumN\phi_{i}^{2}(t)-N\right]\right\},   
\end{eqnarray}  
where the $l$ and $h$ are sources and the last two terms in the
exponent are due to the initial condition and spherical constraint,
respectively. The action $L[\phi,\hfield]$ is given by
\begin{eqnarray}
  L[\phi,\hfield] = -i\sumN \intfin dt\;\hfield_{i}(t)&
  &\left[\Gamma_{0}^{-1}\partial_{t}\phi_{i}(t)-iT\hfield_{i}(t) 
    \right.\nonumber\\*
   & &
    \left.-\frac{p}{p!}\sum_{i_{1}...i_{p-1}}\couplei\fieldls\right]. 
\end{eqnarray}

We now average the generating functional over the disorder in the
system.  The action contains only one term with an explicit dependence
on the disorder, which we call $L_{J}[\phi,\hat{\phi}]$,
\begin {equation}
  L_{J}[\phi,\hfield]=i\frac{p}{p!}\sump \intfin dt
  \couple\hfield_{i_{1}}(t)\phi_{i_{2}}(t)\cdots\phi_{i_{p}}(t). 
\end{equation}
We now compute the part of the generating functional affected by
disorder averaging,
\begin{equation}
  \zJ = \int DJ P\{J\} \exp[L_{J}],          
\end{equation}
with $DJ\equiv\prod\limits_{i_{1}<\cdots<i_{p}}d\couple$.
Therefore, after integrating over the disorder we have $\zJ$ given by 
\begin{eqnarray}
  \zJ = \exp& &\left[-\frac{p^{2}}{p!} \sump\invcouple \intfin \!\! \intfin
    dtdt' \right.\nonumber\\*  
    & &\left.\times \! \sum\limits_{\ar,\ar'\in \{0,1\}}^{C=1,C'=1}
    \sump\prod\limits_{r=1}^{p}\phi_{i_{r}}^{\ar}(t)\phi_{i_{r}}^{\ar'}(t')\right],
\end{eqnarray}
where we have re-labeled the fields using the definitions
$\phi_{i}^{0}(t)\equiv\hfield_{i}(t)$ and
$\phi_{i}^{1}(t)\equiv\phi_{i}(t)$. The constrained variables $C$ and
$C'$ are given by $C\equiv \sum\limits_{r=1}^{p}(1-\ar)$,
$C'\equiv \sum\limits_{r=1}^{p}(1-\ar')$, and the constraints $C=C'=1$
enforce the condition that for each of the two times $t$ and $t'$,
there is a product of fields, of which only one is a $\hfield$ and all
others are $\phi$ fields.  We are also interested in introducing
two-time fields $Q_{i}^{\alpha,\alpha'}(t_1,t_2)$, physically
associated with two-time correlations and responses. In order to do
this we write the number one in terms of an integral of a product of
delta functions that enforce the condition
$Q_{i}^{\alpha,\alpha'}(t_1,t_2)=\phi_{i}^{\alpha}(t_1)\phi_{i}^{\alpha'}(t_2)$,
\begin{equation}
  \label{eq:delta}
    1=\int DQ\prod\limits_{i,t_1,t_2}
    \prod\limits_{\ai,\ai'\in\{0,1\}}
    \delta\left(Q_{i}^{\ai,\ai'}(t_1,t_2)-\phi_{i}^{\ai}(t_1)\phi_{i}^{\ai'}(t_2)\right). 
\end{equation}
By writing the delta function in exponential form we get
\begin{eqnarray}
  1=\int DQ D\hat{Q} \exp&
  &\left\{i \sum\limits_{i}\! \sum\limits_{\ai,\ai'\in\{0,1\}} 
       \int\int dt_1 dt_2
       \hat{Q}_{i}^{\overline{\ai},\overline{\ai'}}(t_1,t_2) \right. \nonumber\\*
   & &\left.\times
       \left(Q_{i}^{\ai,\ai'}(t_1,t_2)-
       \phi_{i}^{\ai}(t_1)\phi_{i}^{\ai'}(t_2)\right)\right\},
\end{eqnarray}
where we have introduced the auxiliary two-time fields
$\hat{Q}_{i}^{\ai,\ai'}(t_1,t_2)$ and the notation $\overline{0}=1$,
$\overline{1}=0$. We now obtain the noise and disorder averaged generating
functional
\begin{eqnarray}
  \z=\int DQ D\hat{Q} D\p D\pp D\hvp D\hat{N} \exp({\cal S}),
  \nonumber \\* 
            {\cal S}=S_{1}+S_{J}+S_{spin}+S_{ext}+S_{BC}+S_{SC}.
\end{eqnarray}
Here we have written the different terms of the action separately:
\begin{eqnarray}
  S_{1}[Q,\hat{Q},\p,\pp]= &
  &i\sum\limits_{i}\! \sum\limits_{\ai,\ai'\in\{0,1\}} 
  \intfin \!\! \intfin dt_1 dt_2 
  \hat{Q}_{i}^{\overline{\ai},\overline{\ai'}}(t_1,t_2)\nonumber\\*
  & & \times\left(Q_{i}^{\ai,\ai'}(t_1,t_2)-\phi_{i}^{\ai}(t_1)\phi_{i}^{\ai'}(t_2)\right),
\end{eqnarray}

\begin{equation}
  S_{J}[Q]= -\frac{p^{2}}{p!} \sump\invcouple\intfin \!\! \intfin dt_1 dt_2
  \! \sum\limits_{\ar,\ar'\in \{0,1\}}^{C=1,C'=1} 
  \prod\limits_{r=1}^{p}Q_{i_{r}}^{\ar,\ar'}(t_1,t_2),
\end{equation}
  
\begin{eqnarray}
  S_{spin}[\p,\pp] &=& -i \sumN \intfin dt
  \p_{i}(t)\left[\Gamma^{-1}\partial_{t}\pp_{i}(t)+ 
    \gamma_{00}\p_{i}(t)\right]-\frac{p^{2}}{p!} \sump\invcouple
  \nonumber\\* 
  & & \times \intfin \!\! \intfin dt_1 dt_2 g(t_1-t_2) 
  \! \sum\limits_{\ar,\ar'\in \{0,1\}}^{C=1,C'=1} 
  \prod\limits_{r=1}^{p}
  \phi_{i_{r}}^{\ar}(t_1)\phi_{i_{r}}^{\ar'}(t_2), 
\end{eqnarray}

\begin{equation}
  S_{ext}[\p,\pp;l,h] =
  \int_{t_{0}}^{t_{f}}dt \left[l_{i}(t)\p_{i}(t)+ih_{i}(t)\pp_{i}(t)\right], 
\end{equation}
 
\begin{equation}
  S_{BC}[\pp,\hvp;\vph]=i\sumN\hvp_{i}\left[\pp_{i}(t_{0})-\vph_{i}\right], 
\end{equation}

\begin{equation}
  S_{SC}[\pp,\hat{N};N]= i\intfin dt
  \hat{N}(t)\left[\sumN(\pp_{i}(t))^{2}-N\right], 
\end{equation}
and we have $\Gamma=\Gamma_{0}$, $\gamma_{00}=-iT$ and $g(t-t')=0$ at
the start of the RG flow.

\section{Renormalization group analysis}
\label{sec:RG}
\subsection{Renormalization Group Transformation}
We perform a renormalization group analysis on the time variables. For
simplicity we take $t_{0}=0$ and $t_{f}=\infty$ from now on. We focus
on the two-time fields.  First, we introduce a cutoff in the
integration of two-time fields, $\tau_{0}\leq|t_1-t_2|$. We then write
the terms of the action affected by the cutoff:
\begin{eqnarray}
  S_{1}[Q,\hat{Q}]=& & i\sum\limits_{i}\intgg dt_1 dt_2\! \sum\limits_{\ai,\ai'}
  \hat{Q}_{i}^{\overline{\ai},\overline{\ai'}}(t_1,t_2)\nonumber\\*
  & & \times\left(Q_{i}^{\ai,\ai'}(t_1,t_2)-\phi_{i}^{\ai}(t_1)\phi_{i}^{\ai'}(t_2)\right),
\end{eqnarray}

\begin{equation}
  S_{J}[Q]=-\frac{p^{2}}{p!} \sump\invcouple\intgg dt_1 dt_2
  \! \sum\limits_{\ar,\ar'\in \{0,1\}}^{C=1,C'=1} 
  \prod\limits_{r=1}^{p}Q_{i_{r}}^{\ar,\ar'}(t_1,t_2).
\end{equation}
We define fast and slow fields respectively by
$Q_{>i}^{\ai,\ai'}(t_1,t_2) = Q_{i}^{\ai,\ai'}(t_1,t_2)$, for
$\tau_{0}\leq|t_1-t_2|<b\tau_{0}$ and
$Q_{<i}^{\ai,\ai'}(t_1,t_2)=Q_{i}^{\ai,\ai'}(t_1,t_2)$, for
$b\tau_{0}\leq|t_1-t_2|$, with $b>1$. This separation of fast and slow
parts of the fields results in a separation in the terms:
\begin{eqnarray}
  S_{1}[Q,\hat{Q},\p,\pp] &=& S_{1}[Q_{>},\hat{Q}_{>},\p,\pp] +
  S_{1}[Q_{<},\hat{Q}_{<},\p,\pp], \\ 
  S_{J}[Q] &=& S_{J}[Q_{>}]+S_{J}[Q_{<}].
\end{eqnarray}
Next we calculate the integral $I_{>}$ over fast fields. To do this we
use the fact that there are no cross-terms between fast and slow
fields in the integral:
\begin{eqnarray}
  I_{>}=\int DQ_{>} D\hat{Q}_{>} \exp\left\{i\sum\limits_{i}\intft
  dt_1 dt_2\! \sum\limits_{\ai,\ai'} 
  \hat{Q}_{>i}^{\overline{\ai},\overline{\ai'}}(t_1,t_2)\right. \nonumber\\* 
  \left.\times\left(Q_{>i}^{\ai,\ai'}(t_1,t_2)-
  \phi_{i}^{\ai}(t_1)\phi_{i}^{\ai'}(t_2)\right)\right. \nonumber \\* 
   \left. -\frac{p^{2}}{p!}\sump\invcouple\intft
   dt_1dt_2\! \sum\limits_{\ar,\ar'\in \{0,1\}}^{C=1,C'=1} 
   \prod\limits_{r=1}^{p}Q_{>i_{r}}^{\ar,\ar'}(t_1,t_2)\right\}.
\end{eqnarray}
Calculating the integral over the ${Q}_{>}$ fields constitutes undoing the delta
function integral transformation we used to introduce the two-time
fields for the fast modes. Hence,  
\begin{equation}
  I_{>}=\exp\left\{-\frac{p^{2}}{p!}\sump\invcouple\intft dt_1
  dt_2\! \sum\limits_{\ar,\ar'\in \{0,1\}}^{C=1,C'=1} 
  \prod\limits_{r=1}^{p}\phi_{i_{r}}^{\ar}(t_1)\phi_{i_{r}}^{\ar'}(t_2)\right\}.
\end{equation}
Next we re-scale all the one-time and two-time fields, thus restoring
the cutoff to its original value $\tau_0$
\begin{eqnarray}
  & &
  Q_{<i}^{\ai,\ai'}(bt'_1,bt'_2)=b^{\lambda_{\ai\ai'}}{Q'}_{i}^{\ai,\ai'}(t'_1,t'_2),\\ 
  & &
  \hat{Q}_{<i}^{\ai,\ai'}(bt'_1,bt'_2) =
  b^{\hat{\lambda}_{\ai\ai'}}{\hat{Q'}}_{i}^{\ai,\ai'}(t'_1,t'_2),\\  
  & & bt'=t, \\ 
  & & \phi_{i}^{\ai}(bt')=b^{\lambda_{\ai}}{\phi'}_{i}^{\ai}(t'), \\ 
  & & l_{i}(bt')=b^{\lambda_{l}}l'_{i}(t'), \\
  & & h_{i}(bt')=b^{\lambda_{h}}h'_{i}(t'), \\
  & & \hvp_i=b^{\lambda_{\vph}}\hvp'_i.
\end{eqnarray}
From the definition of the two time fields in Eq.~(\ref{eq:delta}) and
the transformations of the fields we have
\begin{equation}
  \lambda_{\ai,\ai'}=\lambda_{\ai}+\lambda_{\ai'}.
  \label{eq:scale2to1}
\end{equation}
By rescaling the fields in the part of the action arising from the
disorder average ($S_{J}$) we get  
\begin{eqnarray}
  S'_{J}[Q]=& &-\frac{p^{2}}{p!}b^{\lambda_{J}}\sump\invcouple
  \int\limits_{\tau_{0}\leq|b(t_1'-t_2')|} dt_1' dt_2' \nonumber\\*
  & &\times\! \sum\limits_{\ar,\ar'\in \{0,1\}}^{C=1,C'=1}
  \sump\prod\limits_{r=1}^{p}{Q'}_{i_{r}}^{\ar,\ar'}(t_1',t_2'). 
\end{eqnarray}
Using the relation between $\lambda_{\alpha,\alpha'}$,
$\lambda_{\alpha}$ and $\lambda_{\alpha'}$ given by
Eq.~(\ref{eq:scale2to1}) together with the constraints $C=C'=1$ we get
\begin{equation} 
  \lambda_{J}\equiv 2\lambda_{0}+2(p-1)\lambda_{1}+2.
  \label{eq:lambdaj}
\end{equation}

The other important term we need to consider is $S_{spin}$. When we
rescale the fields we obtain 
\begin{eqnarray}
  S'_{spin}[\p,\pp]=-i\sumN\int_{0}^{\infty} dt'\np_{i}(t')
  \left[b^{\lambda_{vel}}\Gamma^{-1}\partial_{t'}\npp_{i}(t')+
    b^{\lambda_{T}}\gamma_{00}\np_{i}(t')\right]  
  \\*
  -\frac{p^{2}}{p!}\sump\invcouple \int \!\! \int_{0}^{\infty} dt_1'dt_2'
  b^{\lambda_{J}}g(t_1'-t_2')
  \! \sum\limits_{\ar,\ar'\in \{0,1\}}^{C=1,C'=1} 
  \prod\limits_{r=1}^{p}{\phi'}_{i_{r}}^{\ar}(t_1'){\phi'}_{i_{r}}^{\ar'}(t_2')
  \nonumber
\end{eqnarray}
The result is the following set of flow equations
\begin{eqnarray}
  & & \Gamma^{-1}\rightarrow\Gamma^{'-1}=\Gamma^{-1}b^{\lambda_{vel}}, \\
  & & \gamma_{00}\rightarrow\gamma'_{00}=\gamma_{00}b^{\lambda_{T}}, \\
  & & g(t_1-t_2)\rightarrow g'(t_1'-t_2') =
  b^{\lambda_{J}}\left(g(b(t_1'-t_2'))+{\cal
    C}_{\tau_{0}\leq|bt'_1-bt'_2|<b\tau_{0}}\right),  
\end{eqnarray}
where ${\cal C}_{\cal P}$ is defined by ${\cal C}_{\cal P}=1$ if
${\cal P}$ is true and ${\cal C}_{\cal P}=0$ if ${\cal P}$ is not
true, and
\begin{equation}
  \lambda_{vel}=\lambda_{0}+\lambda_{1},
  \label{eq:lambdavel}
\end{equation}
\begin{equation}
  \lambda_{T}=1+2\lambda_{0}.
  \label{eq:lambdat}
\end{equation}

If we let $b=e^{dl}\cong 1+dl$ then the flow equations for $\Gamma$
and $\gamma_{00}$ can be written 
\begin{eqnarray}
  & & \frac{d\Gamma}{dl}=-\lambda_{vel}\Gamma, \\
  & & \frac{d\gamma_{00}}{dl}=\lambda_{T}\gamma_{00}.
\end{eqnarray}

The terms that are left to consider are the constraint terms $S_1$ and
$S_{SC}$, the boundary condition term $S_{BC}$, and the coupling to
the sources $S_{ext}$. Physically we expect that the constraints
represented by $S_1$ and $S_{SC}$ will still be valid in the long time
limit, and therefore those terms should be marginal. This leads to the
equations $\hat{\lambda}_{\overline{\ai},\overline{\ai'}}= -2
-\lambda_{\ai,\ai'} = -2 -\lambda_{\ai} -\lambda_{\ai'}$ and
$\lambda_N = -1$ respectively. It is not obvious from physical
considerations alone what the behavior of the boundary condition and
the coupling to the sources should be. The exponents $\lambda_{l}$,
$\lambda_{h}$, $\lambda_{\vph}$ control their scaling behavior, and
could in principle be chosen to make the terms marginal, but they do
not play any role in what follows.

In order to determine the nature of the fixed points, we need to choose
values for the scaling exponents $\lambda_{0}$ and $\lambda_{1}$.

\subsection{Choice of Scaling Exponents}
\label{sec:choice_lambda}
\begin{figure}[h]
  \centerline{ \includegraphics[width=10cm]{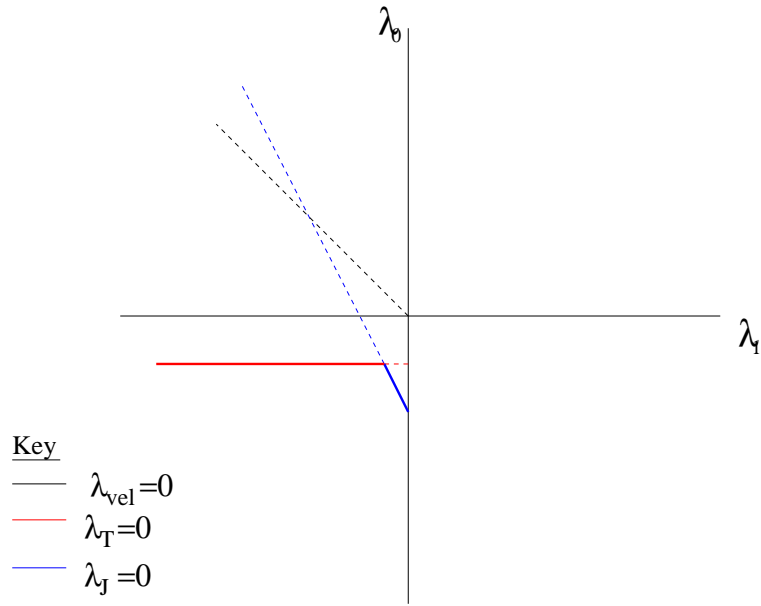} }
  \caption{The figure shows the different lines along which each one
    of the three terms in $S_{spin}$ is marginal for $p=3$. The red
    line corresponds to a marginal coupling to the thermal bath
    ($\lambda_T = 0$), the black line corresponds to a marginal
    time-derivative term ($\lambda_{vel} = 0$), and the blue line
    corresponds to a marginal spin-spin interaction term ($\lambda_{J}
    = 0$).}
  \label{fig:scal_expo}
\end{figure}

In traditional RG calculations one determines an engineering dimension
for the field variable by requiring that the free theory action be
marginal. In our case none of the terms in the action is of the same
form as the gradient squared term that is usually considered to be the
unperturbed part of the action and assumed to be marginal. The only
systematic way to proceed is to consider what happens when each of the
terms in the action is marginal.  We start by noting that the presence
of the spherical constraint $\sumN [\phi_{i}(t)]^{2}=N$ implies that
there is an upper bound on the correlation function $C(t,t')\sim
Q^{11}(t,t')$, and therefore we must have the constraint
$\lambda_1\leq 0$. The case of $\lambda_1=0$ corresponds to freezing
and the strict inequality corresponds to a decaying correlation. The
terms in the action that are of interest for our analysis are the
three terms contained in $S_{spin}$: the spin-spin interaction, the
term containing a time derivative and the term coupling the system to
the thermal bath.  As indicated in Eqs.~(\ref{eq:lambdaj}),
(\ref{eq:lambdavel}) and (\ref{eq:lambdat}), those three terms have
the scaling exponents $\lambda_{J}=2(1+\lambda_{0}+(p-1)\lambda_{1})$,
$\lambda_{vel}=\lambda_{0}+\lambda_{1}$ and
$\lambda_{T}=1+2\lambda_{0}$, respectively.
By considering the cases in which only one of the terms is marginal
we get the results summarized in Fig.~\ref{fig:scal_expo}.

In the case in which the coupling to the thermal bath is marginal, we
have a line $\lambda_{0}=-1/2$ in the ($\lambda_1$, $\lambda_0$)
plane. Considering the constraint $\lambda_{1}\leq 0$ and the values
of $\lambda_{J}$ and $\lambda_{vel}$ we find that there is an interval
on this line, $\lambda_{1}<\frac{-1}{2(p-1)}$, in which both the
spin-spin interactions and the time derivative term are
irrelevant. Since the coupling to the thermal bath is marginal then we
have a family of stable high temperature fixed points.  Second, we
consider the case where the time derivative term is marginal,
corresponding to the line $\lambda_0=-\lambda_1$ in the ($\lambda_1$,
$\lambda_0$) plane. Since we have $\lambda_1\leq 0$, the exponent
$\lambda_{T}$ of Eq.~(\ref{eq:lambdat}) is always positive, i.e. the
coupling to the thermal bath is always a relevant perturbation. Thus
the fixed points that contain only the time derivative term are always
unstable.  We then consider the case where the spin-spin term is
marginal. This happens on the line described by
$\lambda_0=-1-(p-1)\lambda_1$. In the interval
$\frac{-1}{2(p-1)}<\lambda_{1}\leq 0$ the time derivative term and
coupling to the thermal bath are irrelevant. This gives rise to a
family of stable low temperature fixed points. Finally, there is the
special point $\lambda_{1} = \frac{-1}{2(p-1)}, \lambda_{0}=-1/2$, for
which both the coupling to the thermal bath and the spin-spin
interaction are marginal, but the time derivative term is irrelevant,
thus allowing for an additional family of stable fixed points.

The above analysis shows that there is a subset of the ($\lambda_1$,
$\lambda_0$) plane for which a high temperature dynamical fixed point family
is present. The effective generating functional for this fixed point
family is
\begin{eqnarray}
  {\cal Z}_{fp}[l,h;T]=\z_{fp}= \int DQD\hat{Q}D\p D\pp D\hvp
  D\hat{N}\nonumber \\*
  \times\exp\left\{i\sum\limits_{i} \int \!\! \int_{0}^{\infty} dt_1 dt_2
  \! \sum\limits_{\ai,\ai'}\hat{Q}_{i}^{\overline{\ai},\overline{\ai'}}(t_1,t_2)
  \left(Q_{i}^{\ai,\ai'}(t_1,t_2)-\phi_{i}^{\ai}(t_1)\phi_{i}^{\ai'}(t_2)\right)
  \right. \nonumber \\*
  - T \sum\limits_{i=1}^{N} \int_{0}^{\infty}dt
  \left(\phi_{i}^{0}(t)\right)^{2} \nonumber \\*
  \left. + \int_{0}^{\infty}dt
       [l_{i}(t)\p_{i}(t)+ih_{i}(t)\pp_{i}(t)]\right. \nonumber \\*
  \left. +i\sumN\hvp_{i}[\pp_{i}(t_{0})-\vph_{i}]+ i \int_{0}^{\infty}
  dt \hat{N}(t)\left[\sumN(\pp_{i}(t))^{2}-N\right]\right\}. 
\end{eqnarray}
There is another subset of the ($\lambda_1$, $\lambda_0$) plane for
which a low temperature interaction-dominated fixed point family is
present. The effective generating functional for this family of fixed
points is
\begin{eqnarray}
  {\cal Z}_{fp}[l,h;J]=\z_{fp}=\int DQD\hat{Q}D\p D\pp D\hvp
  D\hat{N}\nonumber \\*
  \times\exp\left\{i\sum\limits_{i} \int \!\! \int_{0}^{\infty} dt_1 dt_2
  \! \sum\limits_{\ai,\ai'}\hat{Q}_{i}^{\overline{\ai},\overline{\ai'}}(t_1,t_2)
  \left(Q_{i}^{\ai,\ai'}(t_1,t_2)-\phi_{i}^{\ai}(t_1)\phi_{i}^{\ai'}(t_2)\right)
  \right. \nonumber \\*
  \left. -\frac{p^{2}}{p!}\sump\invcouple \int \!\! \int_{0}^{\infty} dt_1
  dt_2\! \sum\limits_{\ar,\ar'\in \{0,1\}}^{C=1,C'=1}
  \prod\limits_{r=1}^{p}
  Q_{i_{r}}^{\ar,\ar'}(t_1,t_2)\right. \nonumber \\* 
  \left. + \int_{0}^{\infty}dt
       [l_{i}(t)\p_{i}(t)+ih_{i}(t)\pp_{i}(t)]\right. \nonumber \\* 
  \left. +i\sumN\hvp_{i}[\pp_{i}(t_{0})-\vph_{i}]+ 
  i \int_{0}^{\infty} dt
  \hat{N}(t)\left[\sumN(\pp_{i}(t))^{2}-N\right]\right\}. 
\end{eqnarray}
We note that the segment representing stable low temperature fixed
points in the ($\lambda_1$, $\lambda_0$) plane includes the point
$\lambda_1=0$ and $\lambda_0=-1$. This is the only point in the
segment that represents freezing of the correlation, a property of
glasses. 

\section{Time reparametrization symmetry}
\label{sec:TRS}
We now evaluate the effect of a reparametrization $t\rightarrow s(t)$
of the time variable on the stable fixed point generating
functionals. For this purpose we consider a monotonously increasing
function with the boundary conditions $s(0)=0$ and $s(\infty)=\infty$,
which induces the following transformations on the sources,
\begin{eqnarray}
  & & \tilde{l}_{i}(t) =\frac{\partial s}{\partial t}l_{i}(s(t)), \\
  & & \tilde{h}_{i}(t) =h_{i}(s(t)).
\end{eqnarray}
First we consider the effective generating functional of the high
temperature fixed points. We evaluate the fixed point generating
functional of the transformed sources
\begin{eqnarray}
    {\cal Z}_{fp}[\tilde{l},\tilde{h};T]=\int D\tilde{Q}D\tilde{\hat{Q}}D\q D\qq D\tilde{\hvp}
    D\tilde{N} \nonumber \\* 
  \times\exp\left\{i\sum\limits_{i} \int \!\! \int_{0}^{\infty} dt_1 
  dt_2 \! \sum\limits_{\ai,\ai'}
  \tilde{\hat{Q}}_{i}^{\overline{\ai},\overline{\ai'}}(t_1,t_2) 
       \left(\tilde{Q}_{i}^{\ai,\ai'}(t_1,t_2) - 
       \psi_{i}^{\ai}(t_1)\psi_{i}^{\ai'}(t_2)\right)\right. \nonumber
       \\*   
  \left. -T \sum\limits_{i=1}^{N} \int_{0}^{\infty}dt
  \left(\psi_{i}^{0}(t)\right)^{2}\right. \nonumber \\*
  \left. +\int_{0}^{\infty}dt
       [\tilde{l}_{i}(t)\q_{i}(t)+i\tilde{h}_{i}(t)\qq_{i}(t)]\right. \nonumber
       \\*  
  \left. +i\sumN\tilde{\hvp}_{i}[\qq_{i}(t_{0})-\tilde{\vph}_{i}]+i\int_{0}^{\infty}
  dt \tilde{N}(t)\left[\sumN(\qq_{i}(t))^{2}-N\right]\right\}. 
\end{eqnarray}
Here we have used new dummy variables $\psi^{\alpha}$, $\tilde{\hvp}$, $\tilde{\hat{Q}}$, $\tilde{Q}$
and $\tilde{N}$, instead of $\phi^{\alpha}$, $\hvp$, $\hat{Q}$, $Q$ and $\hat{N}$,
respectively, in the functional integral. We now perform the following
change of variables
\begin{eqnarray}
  & & \psi^{\alpha}_{i}(t)= \left(\frac{\partial s}{\partial
    t}\right)^{\overline{\alpha}} \phi^{\alpha}_{i}(s(t)), \\ 
& & \tilde{Q}_{i}^{\alpha,\alpha'}(t,t')=\left(\frac{\partial
    s}{\partial t}\right)^{\overline{\alpha}}\left(\frac{\partial s}
  {\partial t'}\right)^{\overline{\alpha'}}
  Q_{i}^{\alpha,\alpha'}(s(t),s(t')), \\ & &
  \tilde{\hat{Q}}_{i}^{\alpha,\alpha'}(t,t')=\left(\frac{\partial
    s}{\partial t}\right)^{\alpha}\left(\frac{\partial s} {\partial
    t'}\right)^{\alpha'} \hat{Q}_{i}^{\alpha,\alpha'}(s(t),s(t')), \\
  & & \tilde{N}(t)= \frac{\partial s}{\partial t}\hat{N}(s(t)), \\
  & & \tilde{\hvp}= \hvp.
\end{eqnarray}
The change of variables results in Jacobians in the differentials,
\begin{eqnarray}
  & & D\tilde{Q}D\hat{\tilde{Q}}=DQD\hat{Q}{\cal
      J}_{1}\left[\frac{D\tilde{Q}}{DQ}\frac{D\hat{\tilde{Q}}}{D\hat{Q}}\right], \\
  & & D\q D\qq D\tilde{N}=D\p D\pp D\hat{N}{\cal
    J}_{2}\left[\frac{D\q}{D\p}\frac{D\qq}{D\pp}\frac{D\tilde{N}}{D\hat{N}} 
    \right], \\
  & & D\tilde{\hvp}=D\hvp.
\end{eqnarray}
Since the field transformations are linear, the Jacobians depend only
on the reparametrization $s(t)$. Therefore, they are independent of the fields
and sources, and can be taken outside the integral as common factors.

By inserting the values of the transformed sources and dummy variables
back into the fixed point generating functional we obtain,
\begin{eqnarray}\normalsize
  {\cal Z}_{fp}[\tilde{l},\tilde{h};T] = {\cal J}_{1}{\cal J}_{2}\int
  DQ D\hat{Q} D\p D\pp D\hvp D\hat{N} \nonumber \\* 
  \exp\left\{i\sum\limits_{i} \int \!\! \int_{0}^{\infty}
    dt dt'\! \sum\limits_{\ai,\ai'}\left(\frac{\partial s}{\partial
      t}\right)^{\overline{\ai}+\ai} 
       \left(\frac{\partial s}{\partial
         t'}\right)^{\overline{\ai'}+\ai'}
       \hat{Q}_{i}^{\overline{\ai},\overline{\ai'}}(s(t),s(t'))\right.\nonumber\\*  
  \left. -T \int_{0}^{\infty} dt \left(\frac{\partial s}{\partial t}\right)^2
  \left( \phi_{i}^{0}(s(t)) \right)^2 \right. \nonumber \\* 
  \left. +\int_{0}^{\infty} dt \left[ \frac{\partial s}{\partial
    t}l_{i}(s(t))\p_{i}(s(t))+ih_{i}(s(t))\frac{\partial s}{\partial
    t} 
  \pp_{i}(s(t))\right] \right. \nonumber \\* 
  \left. +i\sumN\hvp_{i}[\pp_{i}(s(0))-\vph_{i}]+i\int_{0}^{\infty}
  dt\frac{\partial s}{\partial t} \hat{N}(s(t)) 
  \left[\sumN(\pp_{i}(s(t)))^{2}-N\right]\right\}.
\end{eqnarray}
So then the transformed fixed point generating functional is 
\begin{eqnarray}
    {\cal Z}_{fp}[\tilde{l},\tilde{h};T]={\cal J}_{1}{\cal J}_{2}\int DQD\hat{Q}D\p D\pp
    D\hvp D\hat{N}\nonumber\\* 
    \exp\left\{i\sum\limits_{i} \int \!\! \int_{0}^{\infty}
  dsds'\! \sum\limits_{\ai,\ai'}\hat{Q}_{i}^{\overline{\ai},\overline{\ai'}}(s,s')
  \left(Q_{i}^{\ai,\ai'}(s,s')-
  \phi_{i}^{\ai}(s)\phi_{i}^{\ai'}(s')\right)\right. \nonumber\\*
    \left. -T\int_{0}^{\infty} ds \left(\phi_{i}^{0}(s)\right)^2
    \left( \frac{\partial t}{\partial s} \right)^{-1} \right. \nonumber \\* 
    \left. +\int_{0}^{\infty}ds[l_{i}(s)\p_{i}(s)+ih_{i}(s)\pp_{i}(s)]\right. \nonumber
    \\*  
  \left. +i\sumN\hvp_{i} [\pp_{i}(0)-\vph_{i}]+i\int_{0}^{\infty} ds
  \hat{N}(s)\left[\sumN(\pp_{i}(s))^{2}-N\right]\right\}. 
\end{eqnarray}
Here we have used the fact that $\alpha+\overline{\alpha}=1$.  We
notice that the term describing the coupling to the bath is not
invariant with respect to the transformation $t\rightarrow s(t)$,
except in the trivial case $s(t) = t$. So the high temperature fixed
points are not invariant under reparametrizations of the time
variable.  For the same reason, the fixed point actions containing
both the coupling to the thermal bath and the spin-spin interaction
are not invariant under time reparametrizations.

Finally, we consider the fixed point generating functional for the low
temperature fixed point family.  We evaluate the fixed point
generating functional for the new sources
\begin{eqnarray}
  {\cal Z}_{fp}[\tilde{l},\tilde{h};J]=\int
  D\tilde{Q}D\tilde{\hat{Q}}D\q D\qq D\tilde{\hvp} D\tilde{N}
  \nonumber \\* 
  \times\exp\left\{i\sum\limits_{i} \int \!\! \int_{0}^{\infty} dt_1 
  dt_2 \! \sum\limits_{\ai,\ai'}
  \tilde{\hat{Q}}_{i}^{\overline{\ai},\overline{\ai'}}(t_1,t_2) 
       \left(\tilde{Q}_{i}^{\ai,\ai'}(t_1,t_2) - 
       \psi_{i}^{\ai}(t_1)\psi_{i}^{\ai'}(t_2)\right)\right. \nonumber
       \\*  
  \left. -\frac{p^{2}}{p!}\sump\invcouple \int \!\! \int_{0}^{\infty} dt_1
  dt_2\! \sum\limits_{\ar,\ar'\in \{0,1\}}^{C=1,C'=1} 
       \prod\limits_{r=1}^{p}\tilde{Q}_{i_{r}}^{\ar,\ar'}(t_1,t_2)\right. \nonumber
       \\* 
  \left. +\int_{0}^{\infty}dt [\tilde{l}_{i}(t)\q_{i}(t)+
    i\tilde{h}_{i}(t)\qq_{i}(t)]\right. \nonumber \\*  
  \left. +i\sumN\tilde{\hvp}_{i}[\qq_{i}(t_{0})-\tilde{\vph}_{i}] +
  i\int_{0}^{\infty} dt
  \tilde{N}(t)\left[\sumN(\qq_{i}(t))^{2}-N\right]\right\}. 
\end{eqnarray}
Here we have used the same dummy variables used in the
analysis of high temperature fixed point actions. Perfoming the same 
change of variables we get the Jacobians ${\cal J}_{1}$ and ${\cal J}_{2}$.

By inserting the values of the transformed sources and dummy variables
back into the fixed point generating functional we obtain, 
\begin{eqnarray}\normalsize
    {\cal Z}_{fp}[\tilde{l},\tilde{h};J]={\cal J}_{1}{\cal J}_{2}\int
    DQD\hat{Q}D\p D\pp D\hvp D\hat{N} \nonumber \\* 
    \exp\left\{i\sum\limits_{i} \int \!\! \int_{0}^{\infty}
    dt dt'\! \sum\limits_{\ai,\ai'}\left(\frac{\partial s}{\partial
      t}\right)^{\overline{\ai}+\ai} 
       \left(\frac{\partial s}{\partial
         t'}\right)^{\overline{\ai'}+\ai'}
       \hat{Q}_{i}^{\overline{\ai},\overline{\ai'}}(s(t),s(t'))\right.\nonumber\\*   
 \left.\times\left(Q_{i}^{\ai,\ai'}(s(t),s(t'))-\phi_{i}^{\ai}(s(t))
 \phi_{i}^{\ai'}(s(t'))\right)\right. \nonumber\\* 
   \left.  -\frac{p^{2}}{p!}\sump\invcouple \int \!\! \int_{0}^{\infty}
   dt dt'\! \sum\limits_{\ar,\ar'\in \{0,1\}}^{C=1,C'=1} 
       \prod\limits_{r=1}^{p}\left(\frac{\partial s}{\partial
         t}\right)^{\overline{\ar}} 
       \left(\frac{\partial s}{\partial
         t}\right)^{\overline{\ar'}}Q_{i_{r}}^{\ar,\ar'}(s(t),s(t'))\right. \nonumber
       \\* 
  \left. +\int_{0}^{\infty}dt \left[ \frac{\partial s}{\partial
    t}l_{i}(s(t))\p_{i}(s(t))+ih_{i}(s(t))\frac{\partial s}{\partial
    t} 
        \pp_{i}(s(t))\right] \right. \nonumber \\* 
  \left. +i\sumN\hvp_{i}[\pp_{i}(s(0))-\vph_{i}]+i\int_{0}^{\infty}
  dt\frac{\partial s}{\partial t} \hat{N}(s(t)) 
       \left[\sumN(\pp_{i}(s(t)))^{2}-N\right]\right\}. 
\end{eqnarray}
We now use the fact that $\alpha+\overline{\alpha}=1$ and that the
constraints $C = C' = 1$ ensure that
$\prod\limits_{r=1}^{p}\left(\frac{\partial s}{\partial
  t}\right)^{\overline{\alpha}_{i_r}}\left(\frac{\partial s}{\partial t'}
\right)^{\overline{\alpha}_{i_r}^{'}}=\frac{\partial s}{\partial
  t}\frac{\partial s}{\partial t'}$, to write the transformed fixed
point generating functional
\begin{eqnarray}
  {\cal Z}_{fp}[\tilde{l},\tilde{h};J]={\cal J}_{1}{\cal J}_{2}\int
  DQD\hat{Q}D\p D\pp D\hvp D\hat{N}\nonumber\\*
  \exp\left\{i\sum\limits_{i} \int \!\! \int_{0}^{\infty}
  ds ds'\! \sum\limits_{\ai,\ai'}\hat{Q}_{i}^{\overline{\ai},\overline{\ai'}}(s,s')
  \left(Q_{i}^{\ai,\ai'}(s,s')-
  \phi_{i}^{\ai}(s)\phi_{i}^{\ai'}(s')\right)\right. \nonumber\\*
  \left. -\frac{p^{2}}{p!}\sump\invcouple \int \!\! \int_{0}^{\infty}
  ds ds'\! \sum\limits_{\ar,\ar'\in \{0,1\}}^{C=1,C'=1}
  \prod\limits_{r=1}^{p}Q_{i_{r}}^{\ar,\ar'}(s,s')\right. \nonumber
  \\* \left. +
  \int_{0}^{\infty}ds[l_{i}(s)\p_{i}(s)+ih_{i}(s)\pp_{i}(s)]\right. \nonumber
  \\* \left. +i\sumN\hvp_{i}[\pp_{i}(0)-\vph_{i}]+i\int_{0}^{\infty}
  ds \hat{N}(s)\left[\sumN(\pp_{i}(s))^{2}-N\right]\right\}.
\end{eqnarray}
In other words, we have shown that 
\begin{equation}
  {\cal Z}_{fp}[\tilde{l},\tilde{h};J] = {\cal J}_{1}{\cal J}_{2}
  {\cal Z}_{fp}[l,h;J].
\end{equation}
We know that in the absence of sources, the transformation leaves the
generating functional unchanged. This implies that ${\cal J}_{1}{\cal
  J}_{2}=1$, but since ${\cal J}_1$ and ${\cal J}_2$ are independent
of the values of the sources, then for {\em any} value of the sources
the fixed point generating functional is unchanged by the
transformation, i.e.,
\begin{equation}
   {\cal Z}_{fp}[\tilde{l},\tilde{h}] = {\cal Z}_{fp}[l,h].
\end{equation}
Therefore, for the Langevin dynamics of the p-spin model, the low
temperature long-time fixed point dynamic generating functionals are
symmetric under time reparametrizations.

\section{Discussion and conclusion}
\label{sec:conc}
In our long time renormalization group analysis we have shown that
there are three families of stable fixed point dynamic generating
functionals for the Langevin dynamics of the p-spin model: (i) a
family of high temperature fixed points, which are {\em not \/}
invariant under global reparametrizations of the time variable,
characterized by the presence in the action of the coupling to the
thermal bath and the absence of the spin interaction term; (ii) a
family of low temperature fixed points with actions containing the
spin interaction term but not the coupling to the bath, which are
invariant under global time reparametrizations in the long time limit;
and (iii) a third family of stable fixed points, for which both terms
are present in the action, and thus the action is not invariant under
time reparametrizations.  Since not all of the stable fixed points in the
model are invariant, it is clear that time reparametrization
symmetry is a nontrivial property of the low temperature, interaction
dominated dynamics. It should also be pointed out that in another
interacting spin model, the O(N) ferromagnet, the symmetry is not
present in the asymptotic long time Langevin dynamics, even in the low
temperature case~\cite{chamon}.

The proof of invariance for the low temperature, long time dynamics
only assumes that the couplings $\couple$ are uncorrelated Gaussian
random variables with zero mean, but no condition is imposed on the
variance $\invcouple$ of the couplings, thus allowing them to have an
arbitrary space dependence. In particular, the proof applies to both
short-range and long-range models.  Since some versions of the p-spin
model share many of the main features of structural glass
phenomenology~\cite{Kirkpatrick-etal, Moore-Drossel-prl-2002}, we
expect that analytical tools similar to the ones used here can uncover
the presence of time reparametrization symmetry in models of
structural glass systems.

As discussed in Refs.~\cite{ckcc-prl-2002, cccik-prb-03,
  Castillo-prb-08}, time reparametrization symmetry is a spontaneously
broken symmetry in a glass. The symmetry is broken by correlations and
responses. To illustrate the spontaneous breaking of the symmetry, we
consider the correlation function $C(t_1,t_2)$. If correlations were
invariant under the transformation we would have
$C(t,t')=C(h(t),h(t'))$ for all $t$ and $t'$ and all
reparametrizations and the only way this is possible is when the
correlation function is independent of time. This is not the case in
glasses because the correlation decays with time.  The presence of a
broken continuous symmetry in the absence of long range interactions
or gauge potentials is expected to give rise to Goldstone
modes~\cite{Peskin-Schroeder-quantum}. In the case of the glass
problem, the Goldstone modes should be associated with smoothly
varying local fluctuations $t\rightarrow h_{r}(t)$ in the time
reparametrization~\cite{ckcc-prl-2002, cccik-prb-03,
  Castillo-prb-08}. These fluctuations can be interpreted as
representing local fluctuations of the age of the
sample~\cite{ckcc-prl-2002, cccik-prb-03}. Support for this point of
view comes from simulation results both in the Edwards-Anderson model
of spin glasses~\cite{cccik-prb-03} and in models of structural
glasses~\cite{Castillo-Parsaeian-nphys2007,
  Avila-Parsaeian-Castillo-arxiv-2010}. The kind of analysis performed
in~\cite{Castillo-Parsaeian-nphys2007,
  Avila-Parsaeian-Castillo-arxiv-2010} can in principle be
straightforwardly extended to be applied to particle tracking
experimental data showing dynamical heterogeneities in glassy
colloidal systems~\cite{wks2, Courtland-Weeks-jphys-03}, and in
granular systems~\cite{Durian}.

We conclude by noting that this work hints at the possibility of
analytically proving that time reparametrization symmetry is present
in structural glasses. By investigating the Goldstone modes predicted
as a consequence of the symmetry this may provide an avenue to compute
detailed predictions for probability distributions and correlation
functions that describe the behavior of dynamical heterogeneity.

\ack

We thank L.~Cugliandolo and M.~Kennett for useful discussions, and
particularly C.~Chamon for his help in starting to formulate some of
the ideas presented in Sec.~\ref{sec:choice_lambda}.  This work was
supported in part by DOE under grant DE-FG02-06ER46300, by NSF under
grants PHY99-07949 and PHY05-51164, and by Ohio
University. H.~E.~C. acknowledges the hospitality of the Aspen Center
for Physics and the Kavli Institute for Theoretical Physics, where
parts of this work were performed.

\end{document}